# Proximal nitrogen reduces the fluorescence quantum yield of nitrogen-vacancy centres in diamond

Marco Capelli[a], Lukas Lindner[b], Tingpeng Luo[b], Jan Jeske[b], Hiroshi Abe[c], Shinobu Onoda[c], Takeshi Ohshima[c], Brett Johnson[d], David A. Simpson[e], Alastair Stacey[a,f], Philipp Reineck[a,f], Brant C. Gibson[a,f], Andrew D. Greentree[a,f]

[a] *School of Science, STEM College, RMIT University, Melbourne, VIC 3001, Australia*
[b] *Fraunhofer Institute for Applied Solid State Physics (IAF), Tullastr. 72, 79108 Freiburg, Germany*
[c] *National Institutes for Quantum Science and Technology (QST), 1233 Watanuki, Takasaki, Gunma 370-1292, Japan*
[d] *School of Engineering, STEM College, RMIT University, Melbourne, VIC 3000, Australia*
[e] *School of Physics, University of Melbourne, Melbourne, VIC 3010, Australia*
[f] *ARC Centre of Excellence for Nanoscale Biophotonics, RMIT University, Melbourne, VIC 3001, Australia*

## Abstract

The nitrogen-vacancy colour centre in diamond is emerging as one of the most important solid-state quantum systems. It has applications to fields including high-precision sensing, quantum computing, single photon communication, metrology, nanoscale magnetic imaging and biosensing. For all of these applications, a high quantum yield of emitted photons is desirable. However, diamond samples engineered to have high densities of nitrogen-vacancy centres show levels of brightness varying significantly within single batches, or even within the same sample. Here we show that nearby nitrogen impurities quench emission of nitrogen-vacancy centres via non-radiative transitions, resulting in a reduced fluorescence quantum yield. We monitored the emission properties of nitrogen-vacancy centre ensembles from synthetic diamond samples with different concentrations of nitrogen impurities. While at low nitrogen densities of 1.81 ppm we measured a lifetime of 13.9 ns, we observed a strong reduction in lifetime with increasing nitrogen density. We measure a lifetime as low as 4.4 ns at a nitrogen density of 380 ppm. The change in lifetime matches a reduction in relative fluorescence quantum yield from 77.4 % to 32 % with an increase in nitrogen density from 88 ppm to 380 ppm, respectively. These results will inform the conditions required to optimise the properties of diamond crystals devices based on the fluorescence of nitrogen-vacancy centres. Furthermore, this work provides insights into the origin of inhomogeneities observed in high-density nitrogen-vacancy ensembles within diamonds and nanodiamonds.

## Introduction

The nitrogen-vacancy (NV) centre in diamond is a fluorescent defect consisting of a nitrogen atom neighbouring a vacancy, or missing carbon atom, in the diamond lattice.[1] The negatively charged NV centre (or NV⁻) in diamond has already demonstrated numerous applications,[2] for example as a photostable emitter for biolabeling with nanodiamonds,[3–5] or nanoscale sensor of temperature, magnetic and electric fields.[6–11] These applications take advantage of the NV⁻ high sensing and imaging spatial resolution, as well as operation under ambient conditions.

A key factor in many applications of the NV⁻ centre, especially when using an ensemble of defects, is the emission intensity of the sample. As a first approximation, the emission intensity is proportional to the density of NV⁻ centres, i.e. proportional to the number of emitters. Because of this proportionality, multiple studies investigated methods to improve the creation efficiency of NV⁻ centres in diamond and nanodiamonds.[12–16] However, the fluorescence quantum yield of the NV⁻ centre is commonly thought to be constant and therefore it is not included in the discussion. Despite this, recent studies suggest that the presence of nitrogen defects, as well as other defects, reduces the emission intensity of the NV⁻ centres.[17–19]

Single substitutional nitrogen defects ($N_s$) are ubiquitous in diamond samples with high densities of NV centres therefore, they are of particular relevance in the study of the NV⁻ centre fluorescence. While only the NV⁻ centre has a spin-dependent emission intensity used for quantum sensing applications, the presence of neutrally charged states (NV⁰) contribute to background fluorescence, reducing signal to noise. To maintain a negatively charged state, the NV centre requires an electron donor within the diamond crystal. Nitrogen defects, especially the neutrally charged $N_s^0$ defects, can donate their electron to the NV centre.[20, 21] This process creates pairs of N⁺- NV⁻ in the diamond, as discussed in the literature.[17]

In this study, we show that the fluorescence quantum yield of the NV⁻ centres drop with increasing concentration of $N_s^0$ defects. We investigated a series of HPHT and CVD (chemical vapour deposition) diamond samples whose nitrogen concentrations ranged from 2 ppm up to almost 400 ppm. We characterised the photodynamics–such as emission intensity, lifetime and fluorescence quantum yield–for each sample and as a function of $N_s^0$ concentration. We attribute the change in lifetime and fluorescence quantum yield to the non-radiative tunnelling of the electron within the pair N⁺-NV⁻ while the NV⁻ is in the excited state. Finally, we establish a model curve to predict the change in fluorescence quantum yield from the $N_s^0$ concentration.

The fluorescence quantum yield is defined as the fraction of photons emitted compared to the number of photons absorbed from the excitation source.[22] In a simple 2-level system, the emitter reaches its excited state when it absorbs a photon. The emitter will then return to its ground state producing a photon. In this example, the system has 100% fluorescence quantum yield: for each photon absorbed, a photon is emitted. However, in more complex systems, such as the NV⁻ centre in our study, the relaxation process may involve non-radiative decay pathways, which results in the decrease of the system's fluorescence quantum yield. As an example, the transition from the NV⁻ excited state $m_s = \pm 1$ spin projection to its singlet state is the most well-known non-radiative decay pathway of the NV⁻ centre. The presence of the non-radiative transition to the singlet state–initially observed as a lifetime reduction of the NV⁻ $m_s = \pm 1$ excited state–causes the different spin states of the NV⁻ centre to have different fluorescence quantum yield.[23] Using the emission intensity to distinguish between spin states is the fundamental key that allows most quantum sensing applications with the NV⁻ centre.

However, an overall spin-independent low fluorescence quantum yield has a negative impact on multiple applications of the NV⁻ centres in diamond. For example, a low fluorescence quantum yield hinders the use of nanodiamonds as fluorescent biolabels. NV-containing nanodiamonds already show low brightness per unit mass compared to other biolabels.[24] A further reduction in the nanodiamonds emission intensity, due to the low fluorescence quantum yield, disrupts their viability in bioimaging applications, despite their strong photostability. Additionally, a reduced fluorescence quantum yield has a two-fold impact on applications of the NV⁻ centres in the field of optical sensing. Firstly, the decrease in emission intensity signal reduces the signal to noise ratio–limited by the photon shot noise–of any sensing measurement. Secondly, the non-radiative decay pathway reduces the intensity contrast between the two spin-based emission intensities of the NV⁻ centres, further decreasing the sensitivity of the system.[18]

Previous studies have reported the fluorescence quantum yield of 82 % for shallow implanted NV⁻ centre in single-crystal diamond.[25] However, NV⁻ centres close to the surface of the diamond sample show a reduced fluorescence quantum yield attributed to the dipole interaction with surface defects and group functionalisation.[25, 26] Other studies focus on the reduction of quantum yield caused by irradiation damage in the diamond lattice.[19] Additionally, few studies already observed a change in the emission properties of the NV centres caused by nitrogen defects within the crystal.[17, 18] As aforementioned, the relationship between the NV⁻ centre and nitrogen defects is particularly important due to the prevalence of single nitrogen defects in commonly used diamond samples, such as those created by high-pressure high-temperature (HPHT) synthesis.

HPHT diamond samples are commonly used as starting material for the creation of nanodiamond particles as well as the production of diamond chips with high-density ensembles of NV⁻ centres. However, the HPHT synthesis method naturally incorporates single substitutional nitrogen defects ($N_s^0$) at different concentrations within the same crystal. The difference in nitrogen concentration comes from the presence of growth sectors in each sample.[27, 28] The growth sectors arise from the HPHT synthesis, during which the diamond crystal grows along different growth faces starting from a common seed. The growth faces incorporate defects at different rates and become growth sectors distinguishable by their defect concentrations. In particular, it is typical for the $N_s^0$ concentration to range from 50 ppm up to 400 ppm within different growth sectors in HPHT diamonds. The relatively high concentration of single substitutional nitrogen defects in HPHT diamonds facilitates the creation of high density of NV centres and helps maintain the NV centres in their negatively charged state. However, a low percentage of $N_s^0$ defects are converted into NV centres, and many forms of nitrogen-based defects remain in the crystal lattice after processing. While the interactions between $N_s^0$ defects and the NV⁻ centres have been studied mainly with reference to the change in spin coherence time,[29] we focus on the direct effect of the remaining $N_s^0$ defects on the fluorescence of the NV⁻ centre.

## Instruments & methods

We used single-crystal synthetic diamonds to demonstrate the effect of nitrogen defects on the emission of the NV⁻ centre. We subdivided our diamonds in three distinct batches based on their synthesis process and irradiation parameters. Table 1 summarises the batches we used in this study and their characteristics. The HPHT diamonds in Batch 2 showed growth sectors with different defect concentrations, as discussed above (the growth sectors are visible in the micrographs reported in figure S1 of the Supp. info). Each growth sector within a single HPHT sample shows very distinct optical properties and therefore, we included the different sectors as separate data points.

|  | Supplier | Diamond type | Processing | Additional details |
|---|---|---|---|---|
| Batch 1 | Fraunhofer IAF | CVD IIa | $2\times10^{17}$ cm⁻² electron irradiation 1000°C annealing for 2h | 2 samples |
| Batch 2 | ElementSix | HPHT Ib | $10^{18}$ cm⁻² electron irradiation 900°C annealing for 2h | 2 samples: 3 growth sectors each |
| Batch 3 | Sumitomo Electric | HPHT Ib | $0.5\text{-}5\times10^{18}$ cm⁻² electron irradiation 1000°C annealing for 2h | 4 samples |

*Table 1: Summary of the samples and sectors investigated with associated supplier, diamond type, NV centre creation processing and further details associated with each set. Batch 1 has two CVD-grown samples (type IIa) irradiated at relatively low fluences and used as a benchmark for low-nitrogen behaviour. Batch 2 has two HPHT samples (type Ib), each having multiple growth sectors characterised by different nitrogen concentrations (see figure S1 in the Supp. info). Batch 3 includes four HPHT samples (type Ib) irradiated at different fluences to create different NV centre concentrations starting from samples with similar nitrogen concentrations.*

Batch 1 includes two CVD samples grown at the Fraunhofer IAF, Freiburg. Nitrogen defects were incorporated during the synthesis process of both samples under similar growth conditions. The samples were irradiated with a 2 MeV electron beam to a total fluence of $2\times10^{17}$ cm⁻² at the National Institutes for Quantum Science and Technology (QST), Takasaki, Japan. The samples were annealed at 1000°C for 2 hours to create the NV centres.[14, 30] We used the sample in Batch 1 as a benchmark for the behaviour of an ensemble of NV⁻ centres with low concentrations of $N_s^0$ defects (< 2 ppm). Batch 2 is characterised by samples with different nitrogen concentrations but processed using the same total electron fluence. It includes two HPHT Ib samples from ElementSix. Both samples show the growth sectors caused by HPHT synthesis, as discussed in the introduction. The supplementary information shows micrographs of the samples and their growth sectors visible in transmission under white-light illumination (see figure S1). The samples were irradiated in the same 2 MeV electron beam facility at QST, to a total fluence of $1\times10^{18}$ cm⁻² and annealed afterwards at 900°C for 2 hours. We selected three distinct sectors from each sample for a total of six investigated nitrogen concentrations. Finally, as part of Batch 3, we irradiated four more HPHT Ib samples to different total electron fluences. The samples in Batch 3 were cut from a few larger HPHT samples from Sumitomo Electric before processing. The cut samples do not show inhomogeneity from different growth sectors, except for a small corner in the 1×10$^{18}$ cm$^{-2}$ irradiated sample (see figure S1). Batch 3 samples were irradiated with 2 MeV electrons at QST, however, the total irradiation fluence reached was different for each sample. The samples were part of a series of growing fluence starting from 0.5×10$^{18}$ cm$^{-2}$, then 1×10$^{18}$ cm$^{-2}$, 2×10$^{18}$ cm$^{-2}$ and finally 5×10$^{18}$ cm$^{-2}$. All samples in Batch 3 were annealed at 1000°C for 2 hours. Batch 3 allowed us to verify that our observations do not change with varying NV concentration in the sample and, instead, are predominantly linked to the concentration of $N_s^0$ defects.

Initially, we characterised the samples by visible and infrared absorption spectroscopy to determine the concentration of NV$^-$ centres and nitrogen defects, respectively. We performed visible absorption spectroscopy using a CRAIC Apollo 785 Raman micro-spectrophotometer. The instrument allowed us to measure the absorption signal in transmission from an area of 20×20 µm$^2$ in size. We repeated the absorption measurement three times in different areas within the same sample and growth sector. We used the different areas to calculate the uncertainty in our defects characterisation within each sample and sector. When investigating different sectors, we recorded the measurements in areas far away from the edge between sectors to avoid possible inhomogeneity across the samples' depth. We collected the infrared absorption spectra (FTIR spectroscopy) using a PerkinElmer Spotlight 400. We measured the infrared absorption signal from an area 50×50 µm$^2$ in size and, as before, we recorded three different spectra from each sample/growth sector. The range of wavenumbers in our infrared absorption spectra includes the signatures of different nitrogen-based defects in diamond. We calculated numerically the concentration of $N_s^0$ defects and discussed only qualitatively the presence (or absence) of the other defects in the next paragraph. For both NV$^-$ and $N_s^0$ estimation, we averaged the results between the three areas and used the maximum semi-dispersion of the calculated values as the uncertainty of the result. The suppl. information includes examples of both the visible and infrared absorption spectra (figure S3), together with the exact processing and calculation we used to determine the different concentrations in the samples.

It is important to discuss the charge states of the defects we observed. The estimation of the NV$^-$ concentration is based on the absorption coefficient of the sample at 532 nm.[31] Since the absorption spectra show no trace of the NV$^0$ charge state and the light intensity is not strong enough to ionize the NV$^-$ centre, we can assume that the entirety of the absorption comes from the negatively charged NV centre (NV$^-$). Further analysis of the emission spectra from the samples supports our assumption. Nonetheless, all values of NV$^-$ concentration we report are upper bounds. When discussing nitrogen defects, we only refer to the neutral substitutional single nitrogen defect–or $N_s^0$ defect–commonly found in HPHT Ib diamond crystal.[32] The $N_s^0$ defect is also known in the literature as P1 centre (in electron paramagnetic resonance spectroscopy) or C centre (in infrared spectroscopy). The spectral window used to calculate the concentration of $N_s^0$ defects allows us to discuss the presence of positively charged single nitrogen defects ($N_s^+$)[33] and the double substitutional nitrogen defect (known as A centre in infrared spectroscopy).[34] Our samples show no measurable presence of double substitutional nitrogen, and the $N_s^+$ defect is primarily hidden by the strong absorption of the $N_s^0$ defect in the wide wavenumber range 1050-1350 cm$^{-1}$ (see figure S2 in the suppl. information). The $N_s^+$ defect starts to be identifiable as a peak at 1332 cm$^{-1}$ only on the sample with the largest NV concentration and low $N_s^0$ density. Additionally, we exclude other possible charge states for the single substitutional nitrogen, such as the negative charge state ($N_s^-$) and the double-positive charge states ($N_s^{++}$), since both of them are shown to be unstable under standard experimental conditions.[35]

Finally, the defect concentrations in the CVD samples were determined using different methods. Since the FTIR measurement is not sensitive enough to detect low nitrogen concentrations, we measured the UV-visible absorption spectra of the sample and calculated the concentration of $N_s^0$ defects from their UV absorption band.[36] Similarly, the visible absorption measurement is not able to detect low NV$^-$ concentration, so we determined the NV$^-$ concentration from the emission intensity of the sample. We compared the emission intensity of the samples with unknown NV$^-$ concentration to the emission intensity of a reference sample, using the same optical system. We measured the NV$^-$ concentration of the reference sample using visible absorption spectroscopy as mentioned in the paragraph above. We want to highlight here that this last approach, by our own conclusions discussed further in the paper, is not always reliable and depends on the relative density of $N_s^0$ defects between the unknown and reference samples.

We characterised the emission properties of the samples with a custom-built fluorescence confocal microscope. The samples were excited with a picosecond-pulsed tuneable laser (Fianium whitelase supercontinuum, NKT Photonics). The excitation light spread over a 10 nm band centred at 520 nm at a repetition rate of 10 MHz. A 0.9 NA microscope objective (Plan Apo EPI, Nikon) focused the excitation on the sample as well as collecting the emission from the NV centres. We filtered the emission with a 532 nm long-pass filter to remove the excitation source and with an additional 725 nm long-pass filter to reduce to a minimum the collection of emission from the neutral charge state of the NV centre (NV$^0$). A more detailed study reported in the suppl. information shows that the NV$^0$ contribution in our signal was always less than 0.3 % (figure S6). We fibre-coupled the filtered emission toward an avalanche photodiode detector (APD, AQRH-XX-TR, Excelitas). The intensity signal was sent to a time-tagging correlation card (TimeHarp 260, PicoQuant) for the lifetime measurement. The strength of the emission intensity over time (fluorescence timetrace) was recorded simultaneously with the lifetime measurement, using the open-source QuDi software.[37] We measured both the fluorescence timetrace and the lifetime as a function of excitation power. Example trends are reported in figure S7 of the suppl. information. As we did for the absorption measurements, we collected the timetrace and lifetime curves in three separate areas within each sample and growth sector. The uncertainties in our data reflect the differences between such areas inside a single sample/sector.

## Results & discussion

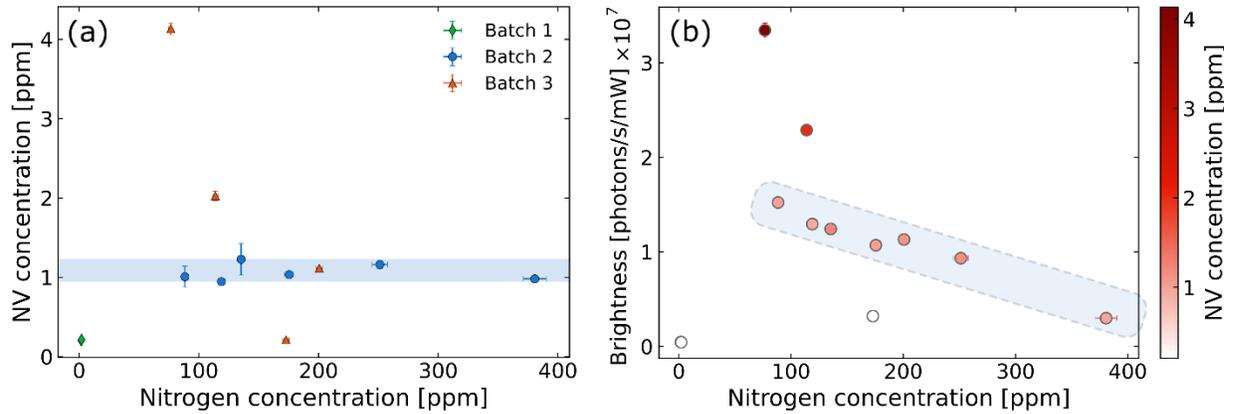

*Figure 1*: (a) Concentration of NV⁻ centres and $N_s^0$ defects for all investigated samples and sectors. The data points are grouped in 3 batches by colour and shape. The green (diamond) data show the CVD samples from Batch 1, which overlap due to their similar growth parameters. Blue (circles) data show the results from the multiple sectors within the HPHT samples in Batch 2. All sectors in Batch 2 were irradiated to the same total fluence, therefore showing similar concentrations of NV centres. The orange (triangles) data from Batch 3 show four different HPHT samples processed with different electron irradiation fluences. The different total fluences cause the creation of varying densities of NV centres, with the highest fluence of $5\times10^{18}$ cm$^{-2}$ corresponding to the sample with more than 4 ppm of NV centres. (b) Brightness distribution of the investigated samples as a function of $N_s^0$ (on the x-axis) and NV⁻ concentration (on the colour axis). The brightness is calculated as the rate at which the emission increases with increasing excitation power (see the suppl. information for more details). The brightness changes with the NV⁻ concentration (represented by the different shades of red), with darker spots corresponding to higher NV⁻ concentrations. The blue shaded areas in both plots highlight the samples irradiated to the same total fluence ($1\times10^{18}$ cm$^{-2}$). In (a) the highlighted samples show a similar NV⁻ concentration (0.944-1.23 ppm) due to the constant irradiation fluence. In (b) the highlighted points show a drop in brightness with increasing nitrogen concentration, even if the NV⁻ concentration remains approximately constant.

Figure 1(a) shows the distribution of $N_s^0$ defects and NV⁻ centre concentration for the samples included in this study. The different batches result in three areas in the scatter plot: (a) in the bottom left corner of the plot we have the CVD-grown samples from Batch 1, which overlap due to their very similar growth parameters. The plot shows they have both low $N_s^0$ and NV⁻ concentration. (b) A series of points positioned along a horizontal line corresponding to an NV⁻ centre concentration of ~1 ppm (highlighted in figure 1(a) by a shaded band). In this area we have the HPHT samples irradiated at a fluence of $10^{18}$ cm$^{-2}$. At a fluence of $10^{18}$ cm$^{-2}$, the concentration of nitrogen impurities in HPHT-synthesised diamond crystals (in the order of 50-400 ppm) is much higher than the density of created vacancies–estimated to be around 10 ppm immediately after irradiation.[38] Therefore, in these samples, the NV concentration is limited by the number of vacancies introduced by the irradiation process. (c) A downward diagonal line, from the top of the plot to its bottom, intersects the 1 ppm horizontal line. This area corresponds to the samples irradiated at different fluences. Starting from the highest point (with an NV⁻ density of 4.13 ± 0.08 ppm) which represents the sample irradiated at a fluence of $5\times10^{18}$ cm$^{-2}$, down to the lowest point of 0.21 ± 0.02 ppm of NV⁻ centres (but with nitrogen concentration of 173 ± 3 ppm) which represents the sample irradiated at a fluence of $0.5\times10^{18}$ cm$^{-2}$.

Figure 1(b) shows how the samples' brightness changes with $N_s^0$ and NV⁻ centre concentration. We calculated the brightness as the rate at which the emission intensity increased with excitation power (see figure S7 in the suppl. information). The NV⁻ concentration in figure 1(b) is expressed via colorscale, from low NV⁻ concentration in white to high NV⁻ concentration in dark red (or dark grey in grayscale). A direct comparison between the plots in figure 1 reveals two main points of discussion:

(1) Considering only the samples at the highest and lowest concentrations of NV⁻ centres (outside the shaded blue areas in both plots) the brightness of the sample directly correlates to the NV⁻ concentration. This result matches our hypothesis. A high concentration of NV⁻ centres corresponds to more NV⁻ centres being excited and thus emitting in the confocal volume of our microscope, which in turn, corresponds to high emission intensity at a fixed excitation intensity. The reverse is true for low concentrations of NV⁻ centres.

(2) The samples with a comparable NV⁻ concentration in figure 1(a) did not show a similar brightness value in figure 1(b). These data points are highlighted in both plots by a shaded blue area. The samples with comparable NV⁻ centre concentration demonstrate a drop of the brightness with increasing concentration of $N_s^0$ defects. The brightness drops by more than a factor of 5 when the nitrogen concentration changes from 88 ± 2 ppm to 380 ± 10 ppm.

Since the drop in brightness cannot be explained by the presence of fewer emitters, we investigated the possible decay pathways from the excited state of the NV⁻ centre using lifetime measurements. We performed lifetime measurements on all samples to understand how the decay rates depend on the concentration of $N_s^0$ defects. The measured lifetime corresponds to the inverse of the sum of all decay rates from the excited state of the NV⁻ centre:

$$\bar{\tau} = \frac{1}{k_{rad} + k_{non-rad} + k_{tunnel}} \qquad (1).$$

We distinguish between 3 possible decay rates: $k_{rad}$ is the radiative rate corresponding with the emission of photons; $k_{tunnel}$ is the non-radiative tunnelling rate within the NV⁻-N⁺ pair, which depends on the distance between the NV⁻ centre and its closest $N_s^0$ defect; $k_{non-rad}$ is any other source of non-radiative decay independent of the nitrogen concentration, such as the transition to the NV⁻ singlet state.

To calculate the average lifetime of each measurement, we fit a stretched exponential to the experimental data (see figure S8 of the suppl. information). Since all our measurements are performed on an ensemble of NV⁻ centres, we expect to measure a distribution of different decay rates.[39] A stretched exponential takes the form of: $I_0 \exp[-(t/\tau_0)^\beta]$. $\tau_0$ represents the time at which the intensity drops by a factor $e$ independently of the parameter $\beta$, and $\beta$ is the dispersion parameter in the range $0 < \beta \leq 1$ that indicates the deviation of the curve from a single exponential (corresponding to $\beta = 1$). $\beta < 1$ corresponds to a wide distribution of lifetimes, with the smaller values of $\beta$ indicating a more spread lifetime distribution. However, $\tau_0$ and $\beta$ do not directly represent the physical parameters of the decaying system.[39] Nonetheless, the fit parameters $\tau_0$ and $\beta$ can be used to calculate the average value of the distribution of lifetimes $\bar{\tau}$ of the ensemble of NV⁻ centres (see the suppl. information for the calculation).

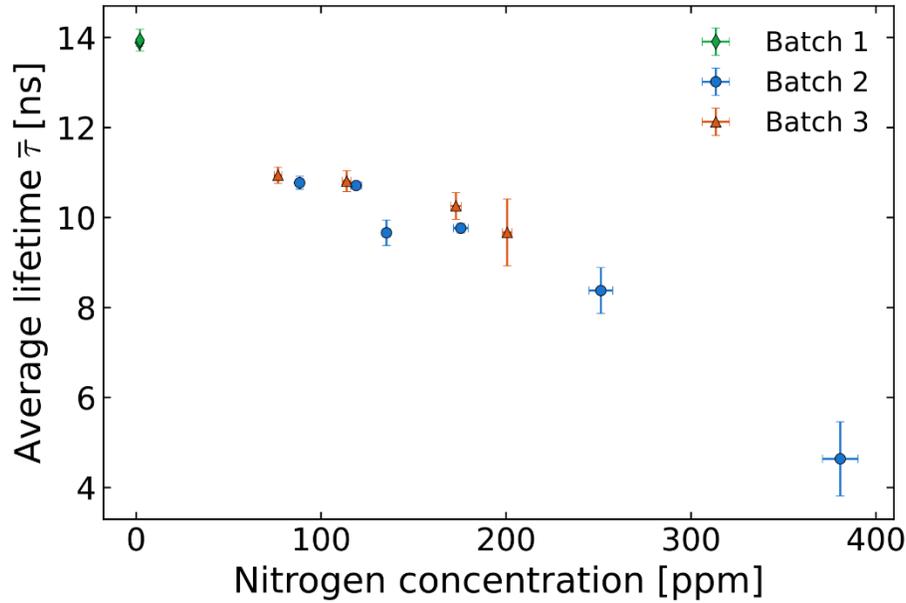

*Figure 2*: *Average fluorescence lifetimes measured from the samples and sectors studied as a function of nitrogen ($N_s^0$) concentration. The scatter plot shows a decrease in the average lifetime with increasing nitrogen concentration.*

We calculated the average fluorescence lifetime for each excitation power. As the excitation power increased, the lifetime plateau toward a constant value. Some examples of the stretched exponential parameters and average lifetimes with excitation power are reported in the suppl. information (figure S9 and A10). We associated each sample with the average lifetime calculated at ~50 µW (~73 kW/cm⁻²), which was the point at which the plateau started. Figure 2 shows that average fluorescence lifetime decreases with increasing $N_s^0$ concentration for all the samples studied in this work. It is important to note that the decreasing trend is independent of the NV⁻ concentration, and only follows the calculated $N_s^0$ concentration.

As shown from equation 1, the observed decrease in the average fluorescence lifetime with $N_s^0$ concentration is explained by the increase of either the radiative rate or the tunnelling rate (or a combination of both). However, an increase in the radiative emission rate would have matched an increase in brightness. We, therefore, attribute the observed decrease in brightness with $N_s^0$ concentration to an increased tunnelling rate. The presence of a nitrogen-dependent non-radiative tunnelling rate matches the current scientific literature, and specifically, the work in [17].

The model proposed in [17] is schematically represented in figure 3(a) and can be briefly described as follows: the negatively charged NV⁻ centre is created by a single nitrogen defect donating its free electron and creating an NV⁻-N⁺ pair. However, the donated electron can tunnel back to the nitrogen donor transforming the NV⁻ into an NV⁰. The NV⁰ will rapidly accept an electron from another (or the same) single nitrogen defect, returning to its NV⁻ state.[40] This latter mechanism also depends on the concentration of $N_s^0$ defects,[41] and prevents us from studying the tunnelling effect through the emission of the $NV^0$ centre. We reported further details on the decrease in $NV^0$ emission with $N_s^0$ concentration in the suppl. information (figures A4 and A6). The entire tunnelling process will result in the relaxation of the NV⁻ to its ground state without the emission of a photon.

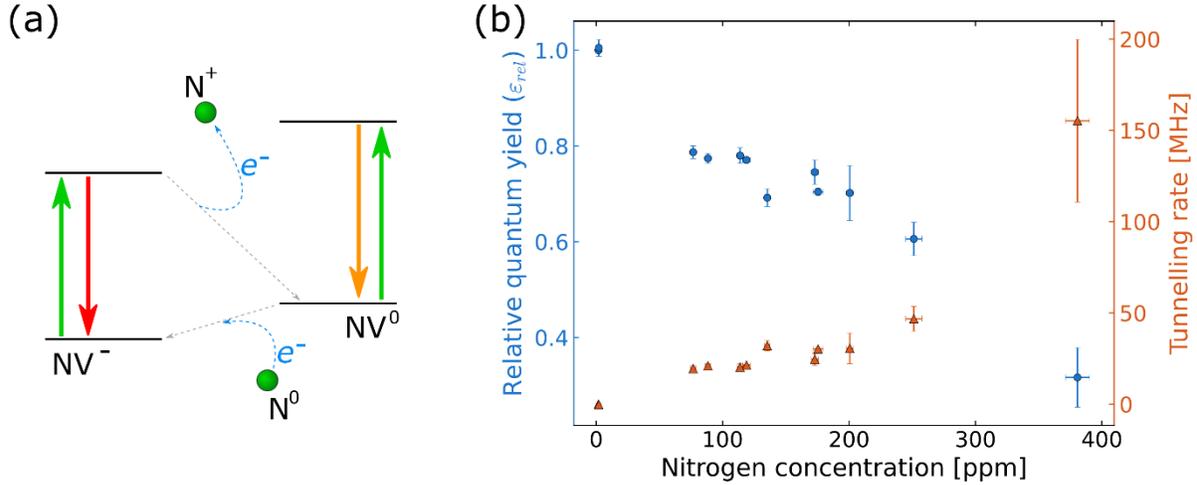

*Figure 3: (a) Schematic of the electron tunnelling between NV centres and nitrogen impurities in the diamond. The solid arrows represent the excitation/emission transitions of the NV centre charge states. The blue curved dashed arrows figuratively indicate the tunnelling of the additional electron to and from the closest nitrogen impurity (represented by the spheres on the edges of the schematic). The grey dashed lines show the equivalent non-radiative transitions between the charge state, which causes the reduced measured lifetime. (b) Relative fluorescence quantum yield (blue dots) and tunnelling rate (orange triangles) calculated from the lifetime measurements as a function of nitrogen concentration. As the nitrogen concentration increases, and hence the average distance between the NV⁻ centres and their closest nitrogen impurity decreases, the tunnelling rate increases. The increase in the non-radiative tunnelling rate corresponds to a decrease in fluorescence quantum yield.*

From theoretical quantum mechanics, the probability of tunnelling increases exponentially with decreasing distance between donor and acceptor. In this work, the average distance between the NV⁻ centre and its closest $N_s^0$ defect represents the distance between donor and acceptor, and it decreases when the density of single nitrogen defects increases. We extracted the tunnelling rate from equation 1, using the calculated average fluorescence lifetime and assuming constant rates for $k_{rad}$ and $k_{non-rad}$. We used the measured lifetime at the lowest $N_s^0$ concentration (1.81 ± 0.01 ppm) in Batch 1 to calculate the constant sum of $k_{rad}$ and $k_{non-rad}$. The decay rate at 1.81 ppm, further referred to as $k_0$, was $k_0$ = 72.0 ± 0.4 MHz which corresponds to a lifetime of $\bar{\tau}$ = 13.90 ± 0.08 ns. Figure 3(b) shows the calculated tunnelling rate $k_{tunnel}$ as a function of $N_s^0$ concentration in each sample (orange triangles).

The presence of any non-radiative decay acts to reduce the fluorescence quantum yield of the emitter, the NV⁻ centre. Although we assume there is no tunnelling at the lowest $N_s^0$ concentration, there are still nitrogen-independent non-radiative rates that decrease the quantum yield below 100%. We define the relative fluorescence quantum yield $\varepsilon_{rel}$ as the ratio between the NV⁻ centre quantum yield and the quantum yield at the lowest $N_s^0$ concentration ($\varepsilon_0$). Figure 3(b) shows the relative quantum yield (blue circles) of the samples as a function of their $N_s^0$ concentration, calculated from the relationship in equation 2.

$$\varepsilon_{rel} = \frac{\varepsilon}{\varepsilon_0} = \frac{k_0}{k_0 + k_{tunnel}} \qquad (2).$$

We observe the highest relative quantum yield of 77.4 % with ~88 ppm of $N_s^0$ defects. However, the relative quantum yield drops to 32 %–less than 1 photon emitted every 3 photons absorbed– with ~380 ppm of $N_s^0$ defects.

Furthermore, we calculated the average distance between an NV⁻ centre and its closest $N_s$ defect from the measured $N_s^0$ concentration using the probability distribution reported in the literature. [42] We used the average distance together with the calculated tunnelling rate to identify the exponential decay expected from theory. Figure 4(a) shows the calculated rates (blue data) with the fitted exponential decay (orange solid line). We run a weighted least-squares minimization method to fit the data, using the inverse square of the data uncertainties as weights. We obtained a maximum tunnelling rate, in the limit of zero-distance, $A$ = 185 ± 87 MHz and a decay constant $\alpha$ = 0.53 ± 0.12 nm⁻¹.

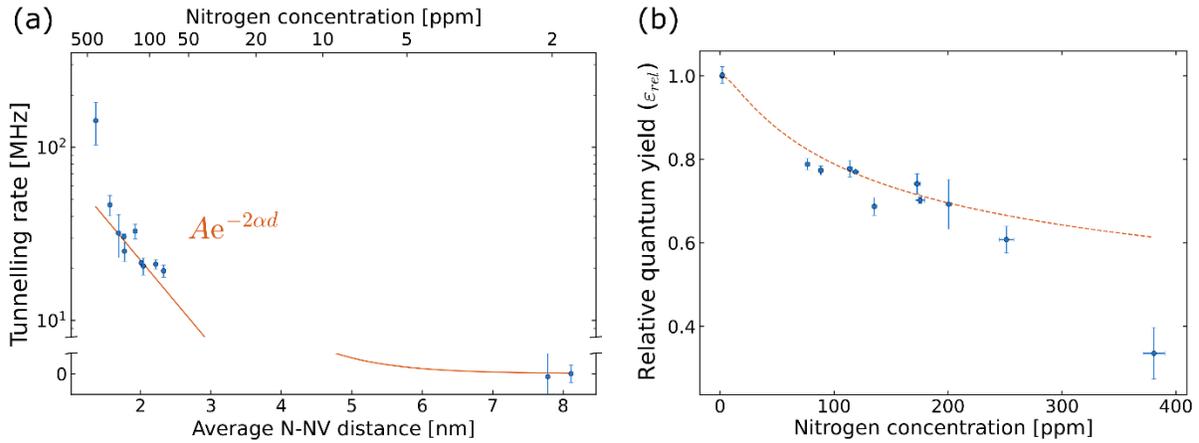

**Figure 4**: *(a) Tunnelling rate as a function of the average distance between an NV centre and its closest single nitrogen defect. The orange solid curve shows the exponential decay fit to the experimental data (blue dots). (b) Relative fluorescence quantum yield data (blue dots) overlayed with the theoretical relative fluorescence quantum yield (dashed orange curve) calculated using the parameters obtained from fitting the data in (a).*

Under the assumption that tunnelling to the closest $N_s^0$ defect is the only, or the predominant, non-radiative mechanism as a function of $N_s^0$ defects, we can write a final equation that links the NV⁻ centre relative fluorescence quantum yield ($\varepsilon_{rel}$) to the $N_s^0$ concentration ($\rho_N$, expressed in nm⁻³) in the diamond sample.

$$\varepsilon_{rel} = \frac{k_0}{k_0 + A\exp[-\alpha_\rho \rho_N^{-1/3}]} \qquad (3).$$

In equation 3, $k_0$ = 72.0 ± 0.4 MHz is the radiative rate we measured from the low-nitrogen data points, $A$ = 185 ± 87 MHz is the same zero-distance tunnelling rate obtained by fitting the tunnelling rate data in figure 4(a), and $\alpha_\rho = \alpha \left(\frac{3}{4\pi}\right)^{1/3} \Gamma\left(\frac{4}{3}\right)$ = 0.294 ± 0.065 nm$^{-1}$ includes the exponential decay constant $\alpha$ and the constant factor that converts the $N_s^0$ concentration (expressed in nm$^{-1}$) in average N-NV distance. Figure 4(b) shows how the curve corresponding to equation 3 follows the experimental data (blue dots). According to the curve in figure 4(b), $N_s^0$ concentrations below 35.5 ppm (6.2×10$^{-3}$ nm$^{-3}$) are required to maintain a relative fluorescence quantum yield above 90%.

Our result can be combined with the change in the NV$^-$ centre coherence time caused by $N_s^0$ defects published in the literature.[29] Both coherence time ($T_2$) and fluorescence quantum yield ($\varepsilon$) affect the total sensitivity of the NV$^-$ centre in high-density ensemble ($\eta_{ens}$) applications, as expressed in equation 4:[9, 43]

$$\eta_{ens} \propto \frac{1}{\sqrt{\varepsilon N \, T_2}} \quad (4),$$

where $N$ is the number of NV$^-$ centres. Previous studies discuss a balance between the number of NV$^-$ centres (which increase with increasing nitrogen concentration) and the coherence time $T_2$ (which decrease with increasing nitrogen concentration). We demonstrated that the fluorescence quantum yield of the NV$^-$ centres decreases with the increase in $N_s^0$ defects concentration, adding a further term in equation 4. This effect on the quantum yield compounds with the decrease of coherence time, breaking the balance with the number of NV$^-$ centres. The combined trend will show a conclusive drop in measurement sensitivity caused by the $N_s^0$ defects.

## Conclusions

We demonstrated the effect of neutral single substitutional nitrogen ($N_s^0$) defects on the fluorescence quantum yield of NV$^-$ centres. $N_s^0$ defects are the predominant defects in many HPHT single-crystal diamonds commonly used for sensing applications with ensembles of NV$^-$ centres. We report the relative fluorescence quantum yield of the NV$^-$ centre to be 77.4 ± 0.9 % in HPHT diamond samples with 88 ± 2 ppm of $N_s^0$ defects, the lowest density we investigated in HPHT diamonds. The relative fluorescence quantum yield drops to 32 ± 7 % in nitrogen-dense (380 ± 10 ppm) diamond samples. The drop of fluorescence quantum yield with $N_s^0$ concentration is explained by the tunnelling of electrons within the N$^+$-NV$^-$ pair, as previously introduced in the literature. According to our best-fit model, to maintain a relative fluorescence quantum yield above 90% the concentration of $N_s^0$ defects must be limited to less than 35.5 ppm. These figures are valuable for the development of the ideal diamond materials for light-based sensing applications with the NV$^-$ centre: from diamond chips created with high-density ensembles of NV$^-$ centres, to bright nanodiamonds observable under commercial microscopes. Finally, our results emphasize the relevance of nitrogen-doped CVD-grown diamonds with a defined concentration of $N_s^0$ defects. Diamond samples created by HPHT methods, which have often been regarded as a good starting material to create ensembles of $NV^-$ centres, show strong variability in the concentration of $N_s^0$ defects. The variability of $N_s^0$ defects is attributed to growth sectors (as demonstrated by Batch 2 in this study) and impacts the performance of $NV^-$ sensing applications, such as inhomogeneity in single-crystal diamond chips and non-uniform emission properties of nanodiamonds.

## Acknowledgements

This work was performed in part at the RMIT Micro Nano Research Facility (MNRF) in the Victorian Node of the Australian National Fabrication Facility (ANFF). We acknowledge the Analytical Chemistry Team, Technical Service, STEM College at RMIT University for the support in collecting the FTIR measurements. The authors want to acknowledge the support of Dr Brian Yang from the ARC Centre for Nanoscale BioPhotonics laboratories at RMIT University and Dr Neil Manson from the Research School of Physics and Engineering (Australian National University, Canberra) for the useful discussion. M.C. acknowledges funding from the Asian Office of Aerospace Research and Development (AOARD, funding FA2386-18-1-4056). J.J. acknowledges funding from the German federal ministry for education and research, Bundesministerium für Bildung und Forschung (BMBF) under Grant No. 13XP5063. P.R. acknowledges support through an Australian Research Council DECRA Fellowship (grant no. DE200100279) and a RMIT University Vice-Chancellor's Research Fellowship. ADG acknowledges financial support from the Australian Research Council (FT160100357). This work was in part carried out within the framework of QST Internal Research Initiative and the MEXT Quantum Leap Flagship Program (MEXT Q-LEAP, grant number JPMXS0118067395).

# Supplementary information

for the paper

# "Proximal nitrogen reduces the fluorescence quantum yield of nitrogen-vacancy centres in diamond"

**White-light transmission micrographs of the samples**

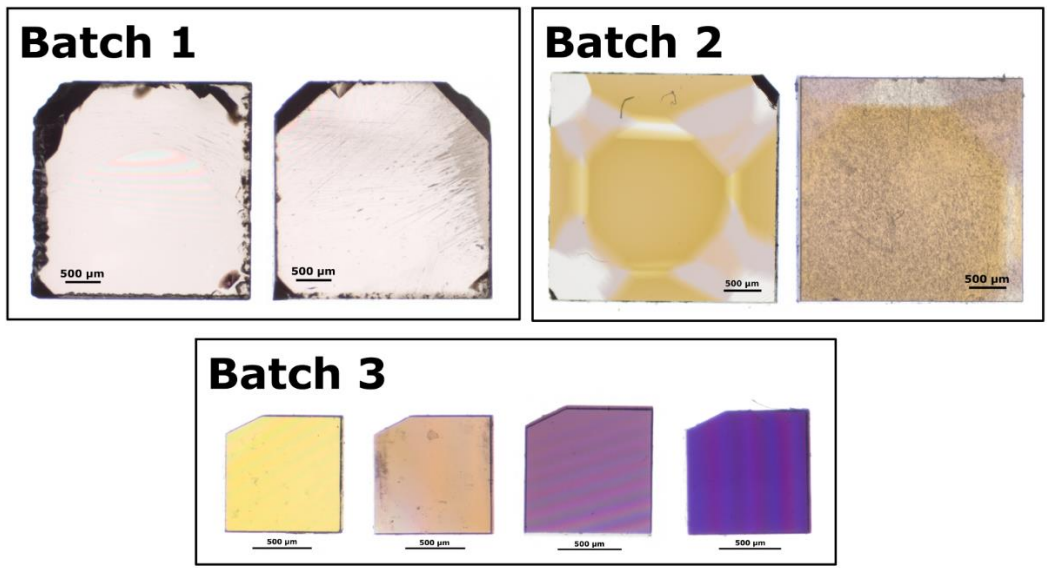

*Figure S1:* Micrographs of the sample included in this study. Batch 1 includes two CVD diamonds. Batch 2 includes 2 HPHT diamonds, each characterised by different growth sectors. Batch 3 include 4 smaller HPHT diamonds irradiated to a different total fluence, from $0.5\times10^{18}$ cm$^{-2}$ on the left side to $5\times10^{18}$ cm$^{-2}$ on the right side. All scale bars are 500 μm.

In this suppl. information, we will report a few example measurements to explain the analysis performed in the main text. Through the full suppl. information, we use the same four samples with different nitrogen concentrations to show our analysis. Figure S2 highlights the samples considered in the suppl. information, using Figure 1(a) of the main text as a reference. The specific NV and nitrogen concentrations of the four samples are reported in Table S1.

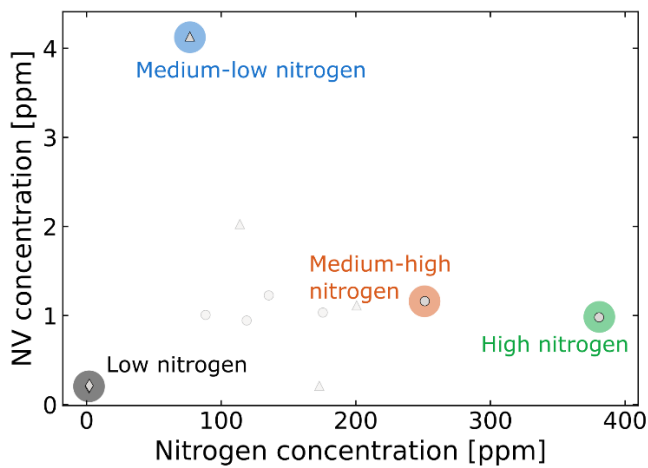

*Figure S2:* Concentration of NV⁻ centres and $N_s^0$ defects for the four samples detailed in the suppl. information. The sample will be referred to as "low nitrogen" sample, "medium-low nitrogen" sample, "medium-high nitrogen" sample, "high nitrogen" sample. The exact NV⁻ and $N_s^0$ concentration values are reported in table S1.

| Sample label name | $N_s^0$ concentration [ppm] | NV⁻ concentration [ppm] |
|:---:|:---:|:---:|
| Low nitrogen | 1.81 | 0.21 |
| Medium-low nitrogen | 77 | 4.13 |
| Medium-high nitrogen | 201 | 1.11 |
| High nitrogen | 381 | 0.98 |

**Table S1:** NV⁻ and $N_s^0$ concentration values for the four samples reported in the suppl. information. While the variation of $N_s^0$ concentration is explicit in the label used for each sample, the NV⁻ concentration is not immediately relatable to the sample name used.

## Calculation of the $N_s^0$ and NV⁻ concentrations

Figure S3 reports the infrared (FTIR) and visible (UV-Vis) absorption spectra of the four reference samples in this suppl. information. The dashed lines indicate the absorption values used to calculate the density of $N_s^0$ defects and NV⁻ centres, from the infrared and visible spectrum respectively.

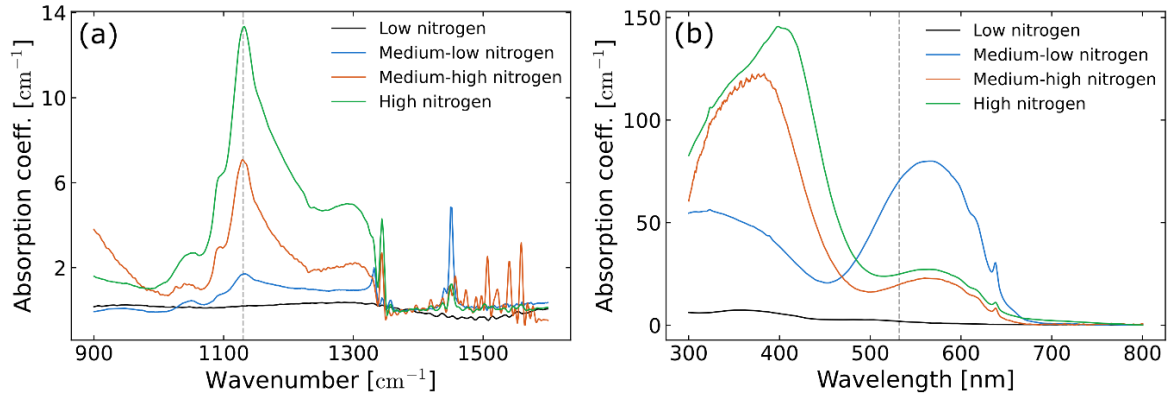

*Figure S3:* Absorption spectra of the suppl. information reference samples in the (a) mid-infrared and (b) UV-visible spectral range. (a) The FTIR measurement is used to calculate the concentration of $N_s^0$ defects, using the absorption at 1130 cm$^{-1}$ highlighted by the dashed vertical line. (b) The UV-Vis measurement shows the absorption of the NV$^-$ centres. The absorption value on the dashed line at 532 nm is used to determine the density of NV$^-$ centres.

The density of $N_s^0$ defects ($\rho_N$) was calculated from the absorption coefficient at 1130 cm$^{-1}$ ($\mu_{1130}$) according to the literature relationship:[32]

$$\rho_N[\text{ppm}] = (25 \pm 2)\, \mu_{1130}\, [\text{cm}^{-1}]$$

The density of NV$^-$ centres ($\rho_{NV}$) was calculated from the absorption coefficient at 532 nm ($\mu_{532}$) and the NV$^-$ cross-section ($\sigma_{NV}$) according to the literature relationship:[31]

$$\rho_{NV}[\text{ppm}] = \rho_C\, [\text{ppm·cm}^3]\, \mu_{532}\, [\text{cm}^{-1}] / \sigma_{NV}\, [\text{cm}^2]$$

The calculations were applied to all samples to obtain figure 1(a) of the main text.

### Contribution of the different NV charge states

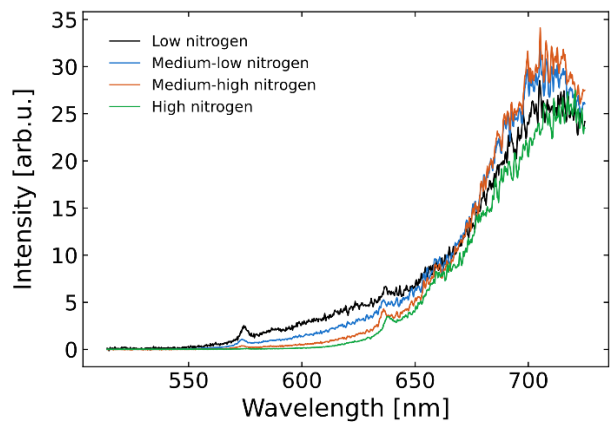

*Figure S4:* Spectral emission of the suppl. information reference samples. The spectra were collected only using a 532 nm long-pass filter to remove the excitation source. The zero-phonon line (ZPL) of the NV$^0$ centre is visible at 575 nm, while the NV$^-$ ZPL is visible at 638 nm. The different samples show that at the same excitation power (~50 µW) the contribution of NV$^0$ emission decreases as the nitrogen concentration increases.

Figure S4 shows the emission of the different NV centre charge states. Using the same excitation power (~50 µW) and normalising the spectrum to the total emission (area under the curve) we notice the change in charge state contribution in the different samples. The $NV^0$ emission is visible in the sample with low nitrogen concentration but decreases as the nitrogen concentration increases. A higher nitrogen concentration corresponds to a larger number of $N_s^0$ defects which act as donors to more easily maintain the NV centres in their $NV^-$ charge state.

We quantified the percentage of $NV^0$ emission to the total NV centre emission by separating the spectra with a non-negative matrix factorisation algorithm. Figure S5 shows an example of the low nitrogen sample emission spectrum separated in its $NV^0$ and $NV^-$ components.

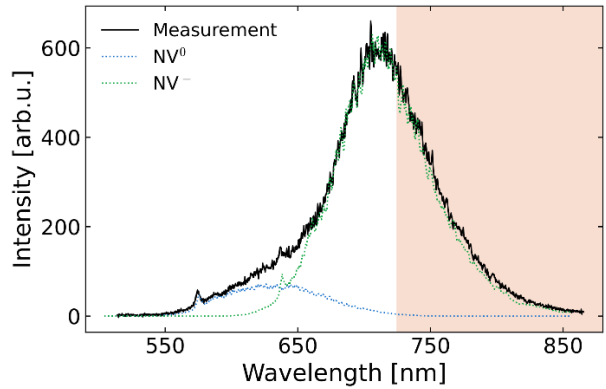

***Figure S5:*** *Emission spectrum of the suppl. information reference sample 'low nitrogen' (black solid line). The non-negative matrix factorisation (NNMF) algorithm separated the $NV^0$ (dotted blue curve) from the $NV^-$ (dotted green curve). The orange shaded area shows the spectral region (725 nm long-pass) used to collect the emission intensity and lifetime in the main text. This filtering window reduced to a minimum the collection of $NV^0$ emission.*

We chose to use a 725 nm long-pass filter to reduce the collection of the $NV^0$ emission in our measurements. However, we collected the full emission spectrum for each sample as a function of excitation power to verify the expected ionisation behaviour and confirm the effect of $N_s^0$ defects on the NV charge state, as shown in figure S6.

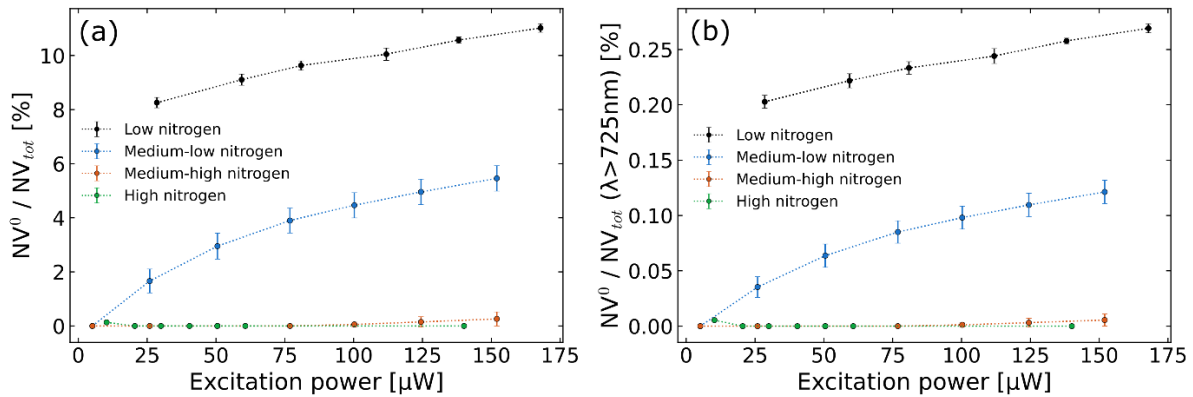

***Figure S6:*** *Percentage of $NV^0$ emission in (a) the total NV emission spectrum and (b) the 725nm long-pass filtered spectrum. The contribution of $NV^0$ emission was calculated from each sample and as a function of excitation power. The percentage of $NV^0$ emission increases with excitation power due to the photo-ionisation of the NV centre. The photo-ionisation is much weaker in the samples with increasing nitrogen concentration.*

Figure S6(b) confirms that the amount of $NV^0$ collected in the measurements reported in the main text (brightness and lifetime) is less than 0.3 %.

**Determination of the samples' brightness**

For all measurements, the excitation power was kept below saturation, in the linear regime of the relationship between emission intensity and excitation power. Figure S7 shows the average emission intensity (calculated from a 60 s timetrace) as a function of excitation power (directly measured with a Thorlabs power meter). The reported brightness (figure 1(b) in the main text) was calculated from the slope of the fitted linear relationship in figure S7.

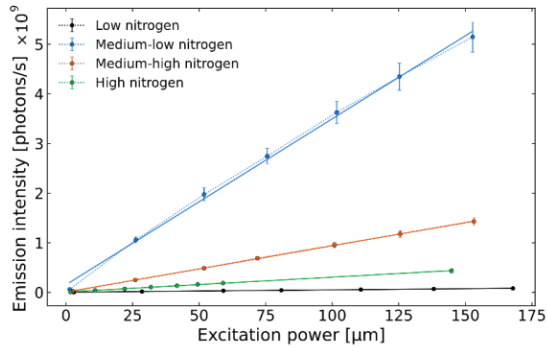

*Figure S7:* Emission curves of the suppl. information reference samples. The solid lines represent linear best-fit to the data. The slope of the best-fit gives the value of brightness reported in the main text. The brightness slopes do not follow the nitrogen concentration trend since the samples shown have different NV concentrations, see table S1. The "low nitrogen" sample (black data) has the lowest density of NV centres as well, resulting in low brightness. The "medium-low nitrogen" sample (blue data) has the highest density of NV centres, due to the highest irradiation fluence, and therefore has the strongest brightness. The remaining "medium-high nitrogen" (orange data) and "high nitrogen" (green data) samples have a similar density of NV centres, thus the effect of nitrogen on the brightness becomes visible.

**Average NV-N distance and distribution of lifetimes**

We calculated the average distance $\langle r \rangle$ between an $NV^-$ centre and its closes $N_s^0$ defect using the formula from the literature:[42]

$$\langle r \rangle = \left(\frac{3}{4\pi\rho_N}\right)^{1/3} \Gamma\left(\frac{4}{3}\right)$$

Where $\rho_N$ is the density of nitrogen defects and $\Gamma$ is the gamma function. The calculated average N-NV distances are used in the dataset shown in figure 4(a) of the main text, with the corresponding $N_s^0$ concentration (in ppm) on the top axis.

Given the tunnelling rate decreases exponentially with the average N-NV distance, at the highest densities of $N_s^0$ defects (short distances) any small variation in the collection volume produces a noticeable difference in the emission lifetime. From ensembles of NV centres (estimated >10000 NVs per collection volume) this variation results in the presence of a continuous distribution of lifetimes collected by each measurement. Therefore, a single exponential decay does not correctly represent the measured lifetime decay curve.

**Average lifetime calculations**

We fit the experimental lifetime curves with a stretched exponential in the form

$$I(t) = I_0 \exp\left[-\left(\frac{t}{\tau_0}\right)^\beta\right]$$

where $\tau_0$ is the timescale corresponding to a decrease in intensity $I_0$ by a factor of $e$, and $\beta$ is a dispersion factor representing how much the curve deviates from a single exponential decay. Figure S8 shows the lifetime measurements from the suppl. information reference samples and their stretched exponential fit.

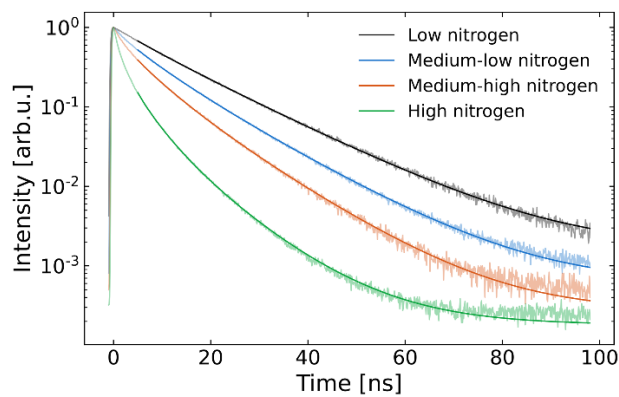

*Figure S8: Lifetime decay curves of the suppl. information reference samples. The solid curves show the stretched exponentials best-fit to the experimental data (shaded).*

Figure S9 show the change in fitting parameters across the reference samples, as a function of excitation power.

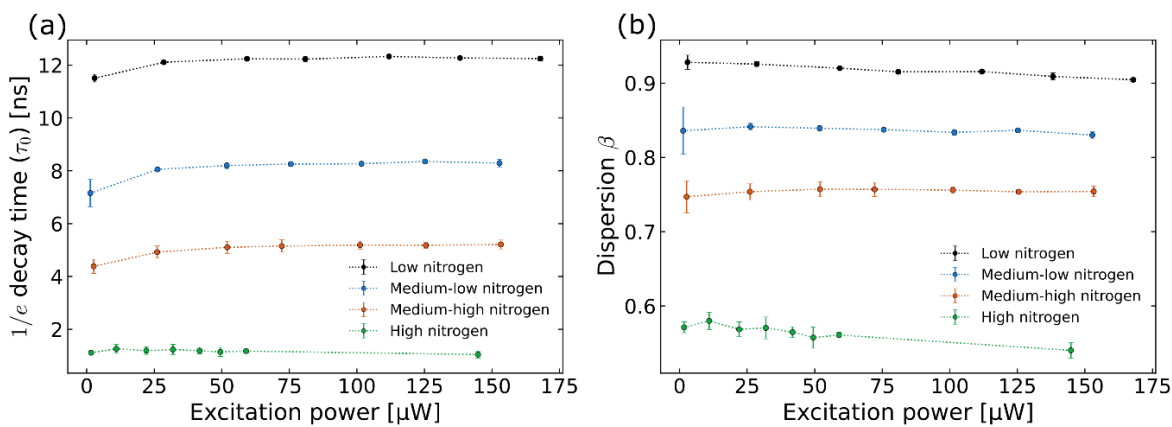

*Figure S9: Stretched exponential fitting parameters (a) decay time $\tau_0$ and (b) dispersion $\beta$ for the suppl. information reference samples as a function of excitation power. While the decay time and dispersion parameters slightly change with excitation power, their values are significantly different for each sample with different nitrogen concentrations. The highest nitrogen concentrations correspond to shorter decay time and smaller dispersion (wide distribution of lifetime values). Low nitrogen concentration approach a single exponential decay (dispersion of 1) and the reported lifetime of a single NV centre in CVD diamond (12 ns).[44]*

The more the dispersion $\beta$ deviates from 1, the wider is the distribution of lifetimes. However, the values of $\tau_0$ and $\beta$ do not represent directly the properties of the lifetime distribution. Instead, the average lifetime $\bar{\tau}$ is calculated from the fitting parameters using the equation:[39]

$$\bar{\tau} = \frac{\Gamma(2/\beta)}{\Gamma(1/\beta)} \tau_0$$

The average lifetimes calculated from the data in figure S9 are shown in figure S10. At low excitation power the average lifetime increases, possibly due to the improved spin-polarisation of the NV⁻ centre. Above ~30 µW the average lifetime starts to approach a constant value, therefore we use the point at 50 µW as the representative value for each sample to build figure 2 of the main text.

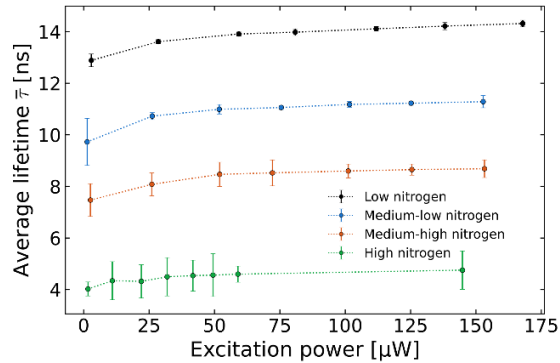

*Figure S10:* Average lifetime for the suppl. information reference samples as a function of excitation power. The average lifetime initially increases with excitation power and then converge to a constant value. The values around 50 µW are used in figure 2 of the main text.